\documentclass{webofc}
\usepackage[varg]{txfonts}   
\newcommand{\beq}[1]{\begin{equation} \label{#1} }
\newcommand{\eeq}   {\end{equation}}

\begin{document}
\title{Angular distribution of fragments in neutron-induced nuclear fission at energies 1-200 MeV: data, theoretical models and relevant problems}
%
%

\author{\firstname{Alexey} \lastname{Barabanov}\inst{1}\fnsep\thanks{\email{a_l_barabanov@mail.ru}} \and
        \firstname{Alexander} \lastname{Vorobyev}\inst{2} \and
        \firstname{Alexei} \lastname{Gagarski}\inst{2} \and
        \firstname{Oleg} \lastname{Shcherbakov}\inst{2} \and
        \firstname{\mbox{Larisa}} \lastname{Vaishnene}\inst{2}
}

\institute{NRC ''Kurchatov Institute'', 123182 Moscow, Russia 
\and
           NRC ''Kurchatov Institute'', B.P.~Konstantinov Petersburg Nuclear Physics Institute, 188300 Gatchina, Russia 
          }

\abstract{%
In recent years, investigations of angular distributions of fragments in neutron-induced nuclear fission have been extended to intermediate energies, up to 200 MeV, as well as to a wide range of target isotopes. Using as an example the latest data obtained by our group for the reaction $^{237}{\rm Np}(n,f)$, we discuss the specific features of fission fragment angular distribution and present a method for their simulation based on the code TALYS. It is shown that a simplified model reasonably describes energy dependence of the angular distribution in the whole range 1--200 MeV. The ways to improve the model are discussed along with the possibilities to use it for obtaining new information on fission and pre-equilibrium processes in neutron-nucleus interaction. We consider also the relevant problems of describing fission fragment angular distributions.
}
\maketitle
\section{Introduction}
\label{intro}

Within the last five years we have studied angular distributions of fragments in neutron induced fission at energies 1--200 MeV for the following isotopes: $^{232}$Th, $^{235}$U, $^{238}$U \cite{Vorobyev_2015}, $^{209}$Bi, $^{233}$U \cite{Vorobyev_2016}, $^{\rm nat}$Pb, $^{239}$Pu \cite{Vorobyev_2018a}, and $^{237}$Np \cite{Vorobyev_2019} (see also \cite{Vorobyev_2017,Vorobyev_2018b}). Measurements were made with the use of the pulsed spallation neutron source and neutron TOF spectrometer GNEIS based on the 1 GeV proton synchrocyclotron in the Petersburg Nuclear Physics Institute \cite{Abrosimov_1985,Shcherbakov_2018}. Similar investigations were performed also by collaborations n$\underline{\phantom{1}}$TOF and NEFFTE for target nuclei $^{232}$Th \cite{Tarrio_2014}, $^{235}$U, $^{238}$U \cite{Leal-Cidoncha_2016} and $^{235}$U \cite{Geppert-Kleinrath_2019}, respectively. These studies are of significance because almost all earlier obtained data on fission fragment angular distributions refer to the energy region below 20 MeV; only in the work \cite{Ryzhov_2005} the angular anisotropy of fragments in neutron induced fission of $^{232}$Th and $^{238}$U was measured up to the energy 100 MeV.

These studies using neutrons of intermediate energies of up to 200 MeV and higher are important for the development of new nuclear technologies, such as nuclear power based on ADS (Accelerator-Driven Systems), transmutation of nuclear waste, radiation testing of materials, nuclear medicine. Besides, the data on fission fragment angular distributions are of interest for nuclear science by stimulating the progress of theoretical models for neutron-nucleus collisions and accompanying processes. Fission is among the most noticeable ones if one uses heavy target nuclei. Moreover, the fission is of special interest due to the fact that a nucleus undergoes fission while being in the equilibrium excited state only, i.e. when the excitation enegy is distributed over all degrees of freedom, including collective ones. That is why at high collision energy, when pre-equilibrium processes play substantial role both in the reaction and subsequent nuclear decay, restrictions on their contributions may be obtained from the observation of fission.

Many other specific features of fission including fission fragment angular anisotropy are related to transition states on barriers first introduced by A.~Bohr \cite{Bohr_1956}. The transition state is characterized by a projection $K$ of the nuclear spin $J$ on the deformation axis which transforms into the direction of fission axis. Thus the quantum number $K$ along with a spin projection $M$ on the $z$ axis of laboratory reference system determines an angular distribution of fission fragments $dw^J_{MK}(\omega)/d\Omega$. Therefore the angular distribution of fragments is among observables, which are the most sensitive to the transition states characteristics, in particular, to the $K$ value. This circumstance resulted in intensive experimental studies of fission fragment angular anisotropy during 2--3 decades after discovery of this phenomenon in the energy region below 20 MeV available at that time (see, e.g., \cite{Vandenbosch_1973}).

In this work we discuss the specific features of fission fragment angular distribution in the wide energy range 1--200 MeV and the possibilities of using them to obtain new information on fission and pre-equilibrium processes in neutron-nucleus interaction. We present a model describing energy dependence of the angular distribution in the whole range 1--200 MeV. Throughout the work we use as an example the data for the reaction $^{237}{\rm Np}(n,f)$. We consider also the relevant problems in this field and perspectives for the development.

\section{Total fission cross section}
\label{sec-1}

Currently, there are a number of computer codes focused on modeling nuclear reactions in a wide energy range, covering the interval of 1--200 MeV; among them the nuclear reaction program TALYS \cite{Koning_2008}. However, TALYS as well as the other known similar programs has a limited functional for describing differential cross sections, in particular, it does not compute differential fission cross section. Let us consider shortly the main features of the TALYS program. It deals only with binary channels, like many other similar codes. Thus, at the first stage of reaction, just after neutron-nucleus collision, only two possibilities are taken into account: either a compound nucleus is formed (let it consists of $Z_0$ protons and $N_0$ neutrons) and then decays into one of the binary channels (fission is among them), or a particle ($\gamma$, $n$, $p$, $d$, $t$, $h$, $\alpha$) and corresponding residual nucleus (which is generally excited) are produced in a direct or pre-equilibrium process. Thus, the cross section of fission at the first stage of the reaction (binary fission in terms of the TALYS Manual) is given by
\beq{t1}
\sigma^{(1)}_f=
\sum_{J\pi}\,
\sigma_{Z_0N_0}(J\pi)\, P_f^{Z_0N_0}(J\pi),
\eeq
where $\sigma_{Z_0N_0}(J\pi)$ and $P_f^{Z_0N_0}(J\pi)$ are the population cross section and the fission probability (fissibility) of the compound state with spin $J$ and parity $\pi$. However, this is only a part of the total fission cross section because any excited residual nucleus formed at any stage of the reaction may undergo fission.

Thus, at the first reaction stage residual nuclei arise either in the primary compound nucleus decay or as the result of a direct or pre-equilibrium process. In the first case, the residual nucleus is assumed to be in an equilibrium excited state (in other words, to be a secondary compound nucleus); it decays at the second reaction stage into a particle and a new residual nucleus or divides into two fragments. In the second case, the residual nucleus is, generally, in a non-equilibrium excited state, therefore, it decays into a particle and a residual nucleus either due to a pre-equilibrium process (this is a multiple pre-equilibrium emission), or after transition to the compound state with an additional possibility to undergo fission. The same is true for third and subsequent reaction stages. Lets $i$ numerates the levels of the nucleus ($Z,N$) with the same spin~$J$ and parity~$\pi$ (in the nucleus ($Z_0,N_0$) the number $i_0$ corresponds to the primary compound state). Then the population cross section $\sigma_{ZN}(J\pi i)$ of a level is a sum of two contributions,
\beq{t1.2}
\sigma_{ZN}(J\pi i)=\sigma^{DPE}_{ZN}(J\pi i)+\sigma^C_{ZN}(J\pi i),
\eeq
where the first one corresponds to the case, when a direct (D) or pre-equilibrium (PE) process preceeds the level population, while the second one --- to the case, when the level is populated as a result of statistical decay of the primary compound nucleus. The total fission cross section is given by
\beq{t2}
\sigma_f=
\sum_{ZNJ\pi i}\,
\sigma_{ZN}(J\pi i)\, P_f^{ZN}(J\pi i),
\eeq
where $P_f^{ZN}(J\pi i)$ is the level fissibility, and summation over $i$ be the integration when the levels belong a continuum (the cross section (\ref{t1}) can be extracted from (\ref{t2}) keeping only the terms corresponding to $Z=Z_0$, $N=N_0$, $i=i_0$ with $\sigma_{Z_0N_0}(J\pi)=\sigma^C_{Z_0N_0}(J\pi i_0)$). We see that according (\ref{t1.2}) and (\ref{t2}) the total fission cross section divides into two parts, $\sigma^{DPE}_f$ and
\beq{t2.2}
\sigma^C_f=
\sum_{ZNJ\pi i}\,
\sigma^C_{ZN}(J\pi i)\, P_f^{ZN}(J\pi i),
\eeq
depending on either we deal with a residual nucleus fission preceeded by a direct or pre-equilibrium process, or with the primary compound nucleus fission or fission of a residual nucleus formed as a result of its statistical decay.

\begin{figure}
\centering
\sidecaption
\includegraphics[width=6.6cm,clip]{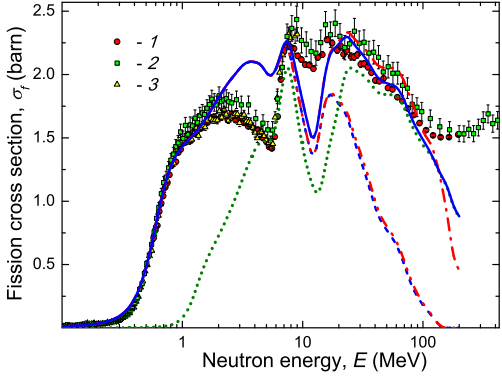}
\caption{Fission cross section for $^{237}$Np versus neutron energy $E$. Experimental data: 1 --- \cite{Shcherbakov_2002}, 2 --- \cite{Paradela_2010}, 3 --- \cite{Diakaki_2016}. Dotted line corresponds to calculation with the default TALYS parameters. Other lines are obtained with the corrected parameters (see text): solid and dot-dashed lines correspond to $\sigma_f$ with and without account for multiple pre-equilibrium emission, dashed and dot-dot-dashed lines correspond to $\sigma^C_f$ with and without account for multiple pre-equilibrium emission.}
\label{f1}       
\end{figure}

In the TALYS program, the fission cross section (\ref{t2}) is calculated using a number of models, in particular, the optical one, as well as for direct and pre-equilibrium processes, level density of excited nuclei, probabilities of radiation transitions, nuclear fission. Each model implies many parameters. One can take the values that are considered as optimal and set by default, or change them. Many default parameters are taken from the RIPL library \cite{Capote_2009}. Since we deal with the reaction $^{237}{\rm Np}(n, f)$, let us  consider the corresponding total fission cross section measured in \cite{Shcherbakov_2002, Paradela_2010, Diakaki_2016}. In the last of these works, the results obtained for the interval 0.1--9 MeV were compared with the calculated values. It turned out that the use of parameters from the RIPL library --- heights and widths of the barriers, transition states, level density at the barriers --- leads to significant discrepancies (see the dotted line in Fig.~\ref{f1} obtained by the TALYS program with the default parameters). But a modification of the listed parameters for the isotopes $^{238}$Np, $^{237}$Np, $^{236}$Np allowed the authors of \cite{Diakaki_2016} to get a good description of the observed fission cross section. We restricted ourselves to correcting the heights and widths of the barriers, as well as the transition states for the mentioned isotopes. Choosing the default option of explicit account of collective enhancement of the level density for nuclei at barriers, we obtained using the TALYS code the total fission cross section as function of the incident neutron energy, which reasonably agrees with the results of  measurements in the range of 0.1--100 MeV. Noticeable discrepancies occur only in the range of 2--16 MeV, but they do not exceed 25-30 \%. The experimental data and the calculated curves are presented in Fig.~\ref{f1}. The calculations were performed with and without account for multiple pre-equilibrium emission (in Ref. \cite{Vorobyev_2019} we have considered only the second option). Both the total fission cross section (\ref{t2}) and its compound component $\sigma^C_f$ (\ref{t2.2}) are shown in Fig. 1. For details about the used corrected parameters for barriers and transition states we refer to \cite{Vorobyev_2019} (transition states correspond to variant 1 from this paper). The calculations of the fission fragment angular anisotropy for the reaction $^{237}{\rm Np}(n,f)$ described below were performed with the same corrected parameters.

\section{Fission fragment angular distribution}
\label{sec-2}

Fission fragment angular distribution is defined via differential fission cross section:
\beq{t3}
W(\theta)\equiv
\frac{dw(\theta)}{d\Omega}=
\frac{1}{\sigma_f}\,\frac{d\sigma_f(\theta)}{d\Omega},
\eeq
where $\theta$ is the angle between the light (for definiteness) fragment momentum and the direction of momentum ${\bf k}$ of incident neutrons, taken for the axis $z$. Notice, that the spin of compound nucleus ${\bf J}={\bf s}+{\bf I}+{\bf l}$ is combined of neutron spin $s=1/2$, target-nucleus spin $I$, and neutron orbital momentum ${\bf l}=[{\bf r}\times {\bf k}]$. When $s$ and $I$ are not oriented, then the spin $J$ is directed predominantly transversely to the $z$ axis because ${\bf l}\perp {\bf k}$. In other words, the compound nucleus states with small projections $M$ of spin $J$ are predominantly populated, with states $M$ and $-M$ are populated with equal probabilities (this type of spin orientation is referred as spin alignment). Spin orientation of fissioning compound state is a necessary condition for angular anisotropy of emitted fragments. When the oriented compound nucleus decays into the particle and residual nucleus, then the spin orientation partially transfers to this nucleus. Thus, the fragments from its fission are emitted also anisotropically with respect to the $z$ axis. The same is true for the subsequent stages of the primary compound nucleus decay. In a similar way, when a direct or pre-equilibrium process occurs at the first reaction stage, then, in principle, the formed residual nucleus also obtains some spin orientation. That is why the fission fragment angular anisotropy may arise when both this nucleus as well as the nuclei resulting from its decay undergoes fission.

Thus, the differential fission cross section (and the angular distribution of fission fragments) depends on the spin orientation of all states (with nonzero spins) of all nuclei forming in the reaction and capable for fission. This orientation, in turn, is determined by the population cross sections $\sigma_{ZN}(J\pi iM)$ for states with particular projections $M$ of spin $J$ on the $z$ axis or, what is equivalent, by the diagonal elements of the spin density matrix
\beq{t4}
\eta_{ZN}^{J\pi i}(M)=\frac{\sigma_{ZN}(J\pi iM)}{\sigma_{ZN}(J\pi i)},\qquad
\sum_M \eta_{ZN}^{J\pi i}(M)=1.
\eeq
These quantities are not calculated in TALYS and other similar programs known to us, therefore, they do not have the option of computing the angular distribution of fragments (as well as the differential fision cross section).

In general, the angular distribution $dw_{ZN}^{J\pi i}(\theta)/d\Omega$ of fragments from fission of a nucleus ($Z$,$N$) at a level ($J$,$\pi$,$i$) is a superposition of the angular distributions
\beq{t5}
\frac{dw^J_{MK}(\theta)}{d\Omega}=
\frac{2J+1}{4\pi}\,|d^J_{MK}(\theta)|^2,
\eeq
corresponding to transition states with definite $J$, $M$ and $K$. In fact, according to A.~Bohr \cite{Bohr_1956}, the spatial part of the wave function of the axially symmetric nucleus at such transition state is proportional to the Wigner function $D^J_{MK}(\omega)=e^{iM\phi}\, d^J_{MK}(\theta)\, e^{iK\psi}$, where Euler angles $\omega=(\phi,\theta,\psi)$ determine the orientation of the reference frame, related to the nucleus, with respect to the laboratory system $(x,y,z)$. Thus, the square of the modulus of the Wigner function determines the distribution in space of the direction of the deformation axis and, consequently, the direction of fission axis. Thus, the superposition has the form
\beq{t6}
\frac{dw_{ZN}^{J\pi i}(\theta)}{d\Omega}=
\sum_M \eta_{ZN}^{J\pi i}(M) \sum_K \rho_{ZN}^{J\pi i}(K)\,\frac{dw^J_{MK}(\theta)}{d\Omega}\,,
\eeq
where the probability distribution over the number $K$ is introduced
\beq{t7}
\rho_{ZN}^{J\pi i}(K)=\frac{P_f^{ZN}(J\pi iK)}{P_f^{ZN}(J\pi i)},\qquad
\sum_K \rho_{ZN}^{J\pi i}(K)=1.
\eeq
Here, $P_f^{ZN}(J\pi iK)$ is the fission probability via the transitional states with the projection $K$ of spin $J$ on the deformation axis (note that variants of expression (\ref{t6}) are sometimes called as the Halpern--Strutinsky formula --- see, e.g., \cite{Krappe_2012}). The distribution (\ref{t7}) is determined by the fission mechanism (in particular, the nucleus may not have axial symmetry on the barrier; in this case, some mixing over $K$ is known to occur). It is important that $\rho^{J\pi}(K)=\rho^{J\pi}(-K)$ in the absence of parity violation effects (they are very small and usually may be neglected). Thus, if the angular distributions (\ref{t6}) normalized to unity are known, then the differential fission cross section can be written as
\beq{t8}
\frac{d\sigma_f(\theta)}{d\Omega}=
\sum_{ZNJ\pi i}
\sigma_{ZN}(J\pi i)\, P_f^{ZN}(J\pi i)\,
\frac{dw_{ZN}^{J\pi i}(\theta)}{d\Omega}
\eeq
and calculated, in effect, by the same scheme as the total fission cross section (\ref{t2}).

In practice, it is convenient to define the spin orientation of the nucleus in the ground or excited state with spin $J$ not in terms of the elements of the density matrix $\eta^J(M)$, but in terms of irreducible components of this matrix or, in other words, in terms of spin-tensors of orientation $\tau_{Q0}(J)$ (see e.g. \cite{Blum_2012}):
\beq{t9}
\tau_{Q0}(J)=\sum_M C^{JM}_{JMQ0}\, \eta^J(M),
\eeq
where $C^{Aa}_{BbDd}$ are Clebsch--Gordan coefficients. For simplicity, the indices $Z$, $N$, $\pi$, $i$ that completely define nuclear levels are omitted here. Restoring them and using the identity
\beq{t10}
(2J+1)\left|D^J_{MK}(\omega)\right|^2=
\sum_Q\, (2Q + 1)\, C^{JM}_{JMQ0}\, C^{JK}_{JKQ0}\, P_Q(\cos\theta)
\eeq
we rewrite (\ref{t6}) as follows
\beq{t11}
\frac{dw_{ZN}^{J\pi i}(\theta)}{d\Omega}=
\frac{1}{4\pi}\,\sum_Q\, (2Q+1)\, \tau^{ZN}_{Q0}(J\pi i)\, \beta^{ZN}_Q(J\pi i)\, P_Q(\cos\theta),
\eeq
where $P_Q(\cos\theta)$ are Legendre polynomials; we call the quantities
\beq{t12}
\beta^{ZN}_Q(J\pi i)=\sum_K C^{JK}_{JKQ0}\, \rho_{ZN}^{J\pi i}(K)
\eeq
the anisotropy parameters. Since $\rho_{ZN}^{J\pi i}(K)=\rho_{ZN}^{J\pi i}(-K)$, they are nonzero only for even values of $Q$. It is also significant that if the distribution $\rho_{ZN}^{J\pi i}(K)$ (or $\eta_{ZN}^{J\pi i}(M)$) is smooth, then the parameters $\beta^{ZN}_Q(J\pi i)$ (or spin-tensors $\tau^{ZN}_{Q0}(J\pi i)$) quickly decrease with increasing of $Q$.

Thus, substituting (\ref{t11}) to (\ref{t8}), we get for the differential fission cross section
\beq{t13}
\frac{d\sigma_f(\theta)}{d\Omega}=
\frac{1}{4\pi} \sum_{Q=0,2,4,\,\ldots} \sigma_{fQ}\, P_Q(\cos\theta)\,,
\eeq
where
\beq{t14}
\sigma_{fQ}=(2Q+1)\sum_{ZNJ\pi i}
\sigma_{ZN}(J\pi i)\, P_f^{ZN}(J\pi i)\,
\tau^{ZN}_{Q0}(J\pi i)\,\beta^{ZN}_Q(J\pi i),
\eeq
with $\sigma_{f0}=\sigma_f$. According to (\ref{t1.2}), the differential fission cross section (\ref{t13}), (\ref{t14}), like the total cross section, devides into the sum of $d\sigma^{DPE}_f/d\Omega$ and $d\sigma^C_f/d\Omega$. As for the angular distribution of the fragments (\ref{t3}), it also takes the form of a series over Legendre polynomials:
\beq{t15}
W(\theta)=
\frac{1}{4\pi}\left(1+\sum_{Q=2,4,\,\ldots} A_Q P_Q(\cos\theta)\right),\qquad
A_Q=\frac{\sigma_{fQ}}{\sigma_f}\,.
\eeq
The convenience of these formulae is determined by the fact that in practice almost always at least one of the distributions, either over $M$ or over $K$, is smooth, so that $\sigma_{fQ}$ and $A_Q$ decrease rapidly with the growth of $Q$. Accordingly, with rare exceptions, the shape of the angular distribution $W(\theta)$ is mainly determined by the single parameter $A_2$.

\begin{figure}
\centering
\sidecaption
\includegraphics[width=6.6cm,clip]{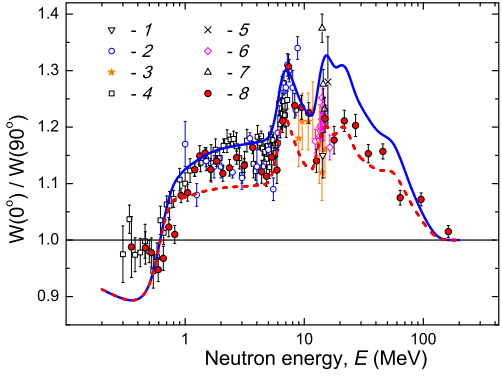}
\caption{Fission fragment angular anisotropy for $^{237}$Np versus neutron energy $E$. Experimental data: 1 --- \cite{Brolley_1954}, 2 --- \cite{Simmons_1960}, 3 --- \cite{Leachman_1965}, 4 --- \cite{Shpak_1971}, 5 --- \cite{Iyer_1970}, 6 --- \cite{Androsenko_1982}, 7 --- \cite{Ouichaoui_1988}, 8 ---\cite {Vorobyev_2019}. The solid and dashed lines correspond to $\hbar^2/J_{\rm eff}=0.022$ and 0.012~MeV.}
\label{f2}       
\end{figure}

It has already been said that we have studied the angular distributions $W(\theta)$ of fragments in the fission of a number of isotopes by neutrons with energies of 1--200 MeV. A detailed description of measuring technique and methods of data processing is presented in \cite{Vorobyev_2015,Vorobyev_2016,Vorobyev_2018a,Vorobyev_2019}. For all isotopes and at all energies, we have found that only the coefficients $A_2$ and $A_4$ are significant in the decomposition (\ref{t15}), and, as expected, the latter is always noticeably smaller than the former. With these coefficients, we have calculated the energy-dependent angular anisotropy
\beq{t16}
\frac{W(0^{\circ})}{W(90^{\circ})}=
\frac{1+A_2+A_4}{1-A_2/2+3A_4/8}
\eeq
for each isotope. In Fig.~\ref{f2} our results are presented on the angular anisotropy of fragments in the reaction $^{237}{\rm Np}(n,f)$ together with the experimental values previously obtained for energies up to 20 MeV \cite{Brolley_1954,Simmons_1960,Leachman_1965,Shpak_1971,Iyer_1970,Androsenko_1982,Ouichaoui_1988}. Note that in some of these works the ratio $W(0^{\circ})/W(90^{\circ})$ had been measured directly. From Fig.~\ref{f2} it can be seen that our data are generally in a good agreement with the results of other authors, the data in the region above 20 MeV are obtained for the first time.

The next section is devoted to our calculations of the fission fragment angular distribution (\ref{t15}), i.e. the coefficients $A_Q$ (as expected, they decrease rapidly with the growth of $Q$). Then, we calculate the angular anisotropy $W(0^{\circ})/W(90^{\circ})$ and compare it with the observed one. In fact, we thus compare the calculated and observed dominant coefficients $A_2$, neglecting the contributions to the angular distribution related to $Q=4$ and above. This is justified at the initial stage of the analysis of angular distributions of fragments. In the following papers we are about to analyze the coefficients $A_4$.

\section{Model for angular distribution of fragments}
\label{sec-3}

To calculate the coefficients $A_Q$ (\ref{t15}), (\ref{t14}) one needs to know the spin-tensors of orientation (\ref{t9}) and the anisotropy parameters (\ref{t12}) for all levels of all fissioning nuclei formed in the reaction. As for the spin orientation of the initial compound nuclei, it can be found in the framework of the optical model (see, e.g., \cite{Vandenbosch_1973}; a general expression for the corresponding spin-tensors is given in \cite{Barabanov_1986}). Transfer of orientation from a decaying nucleus with spin $J_1$ to a residual nucleus with spin $J_2$ in the act of emitting of a light particle with a total angular momentum $j$ is described by
\beq{t17}
\tau_{Q0}(J_2)=\sqrt{(2J_1+1)(2J_2+1)}\,\, W(LJ_2J_1Q,J_1J_2)\, \tau_{Q0}(J_1),
\eeq
where $W(abcd,ef)$ is Racah function (for the case of gamma transition of multipolarity $L=j$ an expression of this type, presumably, had been first obtained in \cite{Cox_1953}). Similarly, spin-tensors of orientation of residual nuclei formed at the first stage of the reaction as a result of a direct or pre-equilibrium process can be calculated in the framework of quantum models of these processes.

Note, however, that the calculated cross-section $\sigma^C_f$ and the observed angular anisotropy of the fission fragments $W(0^{\circ})/W(90^{\circ})$ decrease similarly at $E>20$~MeV (see Figs.~\ref{f1} and~\ref{f2}). Thus, at high energies the angular anisotropy seems be mainly related to the decay of primary compound nucleus. This is so indeed, first, due to the fact that at high collision energy and, consequently, high orbital momentum of incident neutron, the primary compound nucleus possesses a large and well aligned spin $J$.  Second, there is a difference in angular momenta carrying away by particles emitted in statistical decay, on the one hand, and direct and pre-equilibrium processes, on the other hand. In the former case the particles have relatively low energies, of the scale of nuclear temperature, and therefore low angular momenta, while in the later one the particles are usually emitted with high energies and, accordingly, with large angular momenta. Thus, the spin orientation should remain well pronounced for residual nucleus resulting from statistical decay, while there are no grounds for expectations of noticeable spin orientation for residual nuclei resulting from direct and pre-equilibrium processes. Therefore, we take as an assuption that the DPE component of the differential fission cross section is purely isotropic. Thus, instead of (\ref{t13}), (\ref{t14}) we get:
\beq{t18}
\frac{d\sigma_f(\theta)}{d\Omega}=\frac{\sigma_f}{4\pi}+
\frac{1}{4\pi} \sum_{Q=2,4,\,\ldots} \sigma^C_{fQ}\, P_Q(\cos\theta)\,,
\eeq
where the quantities
\beq{t19}
\sigma^C_{fQ}=(2Q+1)\sum_{ZNJ\pi i}
\sigma^C_{ZN}(J\pi i)\, P_f^{ZN}(J\pi i)\,
\tau^{CZN}_{Q0}(J\pi i)\,\beta^{ZN}_Q(J\pi i)
\eeq
are determined by the population cross sections $\sigma^C_{ZN}(J\pi i)$ and spin-tensors of orientation $\tau^{CZN}_{Q0}(J\pi i)$ for states of either primary compound nucleus or residual nuclei arising at its statistical decay.

Besides the spin-tensors of orientation, the anisotropy parameters $\beta^{ZN}_Q(J\pi)$ (\ref{t12}) are to be calculated. As the starting point, we restrict ourselves by introducing a minimum number of additional parameters, and take the distributions $\rho_{ZN}^{J\pi i} (K)$ to be the same for all $Z$, $N$, $J$, $\pi$, and $i$. Above the barrier we use the statistical distributions, $\rho_{ZN}^{J\pi i}(K)\sim e^{-K^2/2K_0^2}$, where $K_0^2=J_{\rm eff} T/\hbar^2$ is determined by the temperature $T$ of the nucleus on the barrier and an effective moment of inertia $J_{\rm eff}$ (see \cite{Vandenbosch_1973}), while below the barrier we assume $\rho_{ZN}^{J\pi i}(K)\sim e^{-\alpha (|K|-K_1)^2}$, where $\alpha$ is a fixed parameter and $K_1$ is the spin projection for the dominant transition state. Note that at high energies of incident neutrons, the excitation energy of a fissioning nucleus, as a rule, sizably exceed the barrier height, and therefore, generally, the type of distribution over $K$ for energies below the barrier is not of importance. Only at low neutron energies, at the sub-barrier fission of the compound nucleus $^{238}$Np, the value of $K_1$ for this isotope essentially determines the type of angular distribution of the fragments; we discuss this point in details below. The choice of $K_1$ for the isotope $^{237}$Np can noticeably affect the angular anisotropy of the fragments in the threshold energy region for the reaction $(n,n^{\prime}f)$, near $E\simeq 7$~MeV. The experimental data reveal a pronounced peak (see Fig.~\ref{f2}), so to maximize the effect the lowest value $K_1=0.5$ is taken. For all other nuclei the value for $K_1$ was assumed to be $1.5$. Additional explanations are presented in \cite{Vorobyev_2019}.

Thus, the quantities (\ref{t19}) for $Q=2,4,\ldots$ can be calculated by the same scheme as the compound component of the total fission cross section (\ref{t2.2}). It was realized as an addition to the TALYS program. 

\section{Results and discussion}
\label{sec-4}

The model described above and the results of first calculations for fission fragment angular distribution in the reaction $^{237}{\rm Np}(n,f)$ were previously presented by us in the paper \cite{Vorobyev_2019}. The main result was that even in its simplest form, with the use of a minimum number of additional parameters, the model successfully reproduces the gross structure of the dependence of the angular anisotropy of fragments on energy in a wide energy range of 1--200 MeV. In this paper, we present some additional results, as well as consider in more details relevant problems.

The first is that at this stage we, in fact, neglect the influence of direct and pre-equilibrium processes on angular anisotropy. There are reasons for this, as shown above, but it would be interesting to obtain quantitative estimates. Note that in \cite{Vorobyev_2019} we used a simplification that reduces the role of these processes and hence the uncertainties associated with them, namely: the angular distributions of the fragments were computed in neglecting the multiple pre-equilibrium emission. However, as seen from Fig.~\ref{f1}, the account for multiple pre-equilibrium emission only slightly lowers the fission cross section in the region of 20--100 MeV and practically does not affect the compound component $\sigma^C_f$ of this cross section. Similarly, it turn out that the angular distributions calculated in the framework of the model are scarcely dependent on whether the multiple pre-equilibrium emission is taken into account or not. In this work, the angular distributions are calculated with account for the multiple pre-equilibrium emission.

It has already been mentioned above that in the current simplified version of our model one value is used for the effective moment of inertia on the barrier $J_{\rm eff}$ for all fissioning isotopes. To establish the sensitivity to this parameter, the fission fragment angular anisotropy was calculated for different values of $J_ {\rm eff}$. The results are presented in Fig.~\ref{f2} for two values with $\hbar^2/J_{\rm eff}=0.012$ and 0.022~MeV. These values have been chosen so that to include almost all experimental points above $\sim 1$~MeV into region limited by the corresponding curves. As expected, the anisotropy does not depend on $J_{\rm eff}$ at energies below 1 MeV, where the subbarrier fission occurs (in these calculations it is taken $K_1=0$ for $^{238}$Np), and above 100 MeV, where the observed angular distribution becomes close to isotropic. In reallity, the effective moment of inertia varies from nucleus to nucleus, and this calculation shows the scale of its variation. We plan to include in the model the possibility of taking into account the dependence of $J_{\rm eff}$ on the type of fissioning nuclei. In general, the higher the collision energy, the greater the number of fissioning isotopes contributing to the observed fission cross sections (earlier, the importance of proper account for multi-chance fission was pointed out, in particular, in \cite{Ryzhov_2005}). Thus, the angular anisotropy of fragments seems to be sensitive to systematics describing not only the effective moment of inertia, but also other parameters significantly dependent on $(Z,N)$, such as the parameters of fission barriers and level density.

Special attention is to be paid to the discrepancy between the calculated and observed fission cross-section in the region above 100 MeV --- see Fig.~\ref{f1}. It seems likely the models used in TALYS overestimate the role of pre-equilibrium processes in this area,  and, thus, underestimate the yield of fissioning primary and secondary compound nuclei. In particular, our experimental data (see Fig.~\ref{f2}) show the presence of a small but nonzero anisotropy above 100~MeV, although the calculated values of $W(0^{\circ})/W (90^{\circ})$ in this region practically coincide with the unit. Therefore, there is good reason to believe that further studies of the fission fragment angular anisotropy at intermediate energies will refine the contributions of pre-equilibrium processes to cross sections of different processes, in particular, of fission.

\begin{figure}
\centering
\sidecaption
\includegraphics[width=6.6cm,clip]{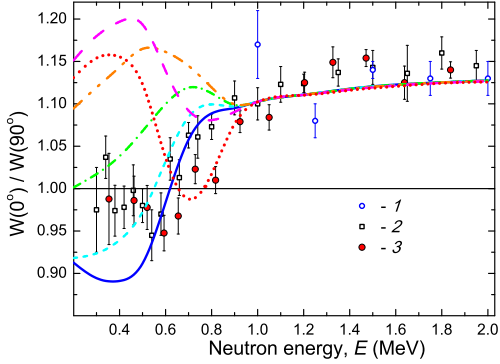}
\caption{Fission fragment angular anisotropy for $^{237}$Np versus neutron energy $E$ for different values of the parameter $K_1$ of the model (see text). Experimental data: 1 --- \cite{Simmons_1960}, 2 --- \cite{Shpak_1971}, 3 ---\cite {Vorobyev_2019}. The solid line corresponds to $K_1=0.0$, the short-dashsed line --- to $K_1=1.0$, the dot-dashed line --- to $K_1=1.5$, the dot-dot-dashsed line --- to $K_1=2.0$, the long-dashed line --- to $K_1=3.0$, the dotted line --- to $K_1=4.0$.}
\label{f3}       
\end{figure}

Finally, let us turn to energies of incident neutrons below 1 MeV. In this region the fission cross-section falls rapidly due to transition from above-barrier to sub-barrier fission. Here one would expect a significant effect of a low-lying transition state with a definite value of $K_1$ on the angular distribution of the fragments. However, the reality seems more complicated as follows from the calculations of angular anisotropy performed for  different values of $K_1$ --- see Fig.~\ref{f3} (note that all the curves at this figure have been obtained with $\hbar^2/J_{\rm eff}=0.017$~MeV). In the region under consideration, especially below 0.5~MeV, neutron $s$- and $p$-waves dominate. Thus, since the nucleus $^{237}$Np has spin and parity $5/2^+$, the compound states $2^+$, $3^+$, $1^-$, $2^-$, $3^-$, $4^-$ are predominantly excited in neutron capture. So the spin projection $K_1$ was taken in the range from 0 to 4. Although the accuracy of the experimental data is not perfect, it can be seen that none of the curves describes them satisfactorily. The most appropriate are the extreme values $K_1$, equal to 0 and 4. Thus, in the sub-barrier energy region our approach to distribution over $K$ does not look suitable. However, do note that the data \cite{Shpak_1971} had been obtained a long time ago, but no successful attempt to interpret them is known. 

\section{Summary}
\label{sec-5}

Angular distributions of fragments in neutron-induced nuclear fission have attracted the attention of researchers for many decades. Substantial progress was achieved in \cite{Ryzhov_2005}, where for the first time not only experimental data for energies noticeably exceeding 20 MeV were obtained, but also an attempt was made to describe the data using modern computer codes on the basis of a model similar to that described in this paper. This model, in particular, was based on the calculations of the population cross sections of nuclear states with the spin projections $M$ on the $z$ axis. The results obtained for the reactions $^{232}$Th ($n,f$) and $^{238}$U ($n,f$) were compared with the experimental data presented in the same paper. However, no further publications of these authors came out, to  unveil the necessary details on the method used or perform calculations for other target nuclei.

In recent years, invetigations have been extended to intermediate energies, up to 200 MeV, as well as to a wide range of target isotopes. The need to systematize new data, extract from them essential information about the fission process as well as the features of neutron-nucleus reactions at intermediate energies prompted us to develope a method for calculating the fission fragment angular distributions. In this paper we present the method based on the TALYS program and show that the results of calculations based on the simplified model are in good agreement with experimental data. It provides a hope that new information about fission barriers, transition states, as well as the role of pre-equilibrium processes in the neutron-nuclear interaction can be obtained by improving and complicating the model.

At this stage, we see the need to complicate our model by including an account for the dependence of parameters such as the effective moment of inertia on the type of fissioning nuclei. It would be useful to develop methods for calculating the spin-tensors of orientation for excited nuclei formed in direct and pre-equilibrium processes. At low energy region, it is necessary to combine in some way an evaluation of distribution of fission probabilities over $K$ with the usually performed calculation of fission transmission coefficients involving account for appropriate transition states. The contribution related to the fourth Legendre polynomial into fission fragment angular distribution calls for further investigation. In general, it seems that nowadays the  experimental data on angular distributions of fragments are little used for improving the current fission model, thus the main problem is to correct this situation.

At last, mention that the developed methods can be used to calculate the angular distributions of $\gamma$-quanta emitted by excited compound nuclei, which, for example, may be useful in the study of the $(n,\gamma f)$ reaction. In addition, all these methods can be extended to reactions initiated by charged particles, such as $(p,f)$ and $(p,\gamma)$. 
\bigskip

Partial support from the RFBR Grant No.~18-02-00571 is greatly acknowledged.


\begin{thebibliography}{88}

\bibitem{Vorobyev_2015} A.S.~Vorobyev, A.M.~Gagarski, O.A.~Shcherbakov, L.A.~Vaishnene, A.L.~Barabanov, JETP~Lett. {\bf 102}, 203 (2015)

\bibitem{Vorobyev_2016} A.S.~Vorobyev, A.M.~Gagarski, O.A.~Shcherbakov, L.A.~Vaishnene, A.L.~Barabanov, JETP~Lett. {\bf 104}, 365 (2016)

\bibitem{Vorobyev_2018a} A.S.~Vorobyev, A.M.~Gagarski, O.A.~Shcherbakov, L.A.~Vaishnene, A.L.~Barabanov, JETP~Lett. {\bf 107}, 521 (2018)

\bibitem{Vorobyev_2019} A.S.~Vorobyev, A.M.~Gagarski, O.A.~Shcherbakov, L.A.~Vaishnene, A.L.~Barabanov, JETP~Lett. {\bf 110(4)}, in print (2019)

\bibitem{Vorobyev_2017} A.S.~Vorobyev, A.M.~Gagarski, O.A.~Shcherbakov, L.A.~Vaishnene, A.L.~Barabanov, EPJ Web Conf. {\bf 146}, 04011 (2017)

\bibitem{Vorobyev_2018b} A.S.~Vorobyev, A.M.~Gagarski, O.A.~Shcherbakov, L.A.~Vaishnene, A.L.~Barabanov, Bull. RAS Phys. {\bf 82}, 1240 (2018)

\bibitem{Abrosimov_1985} N.K.~Abrosimov, G.Z.~Borukhovich, A.B.~Laptev, V.V.~Marchenkov, G.A.~Petrov, O.A.~Shcherbakov, Yu.V.~Tuboltsev, V.I.~Yurchenko, Nucl. Instrum. Methods Phys. Res.~A \textbf{242}, 121 (1985)

\bibitem{Shcherbakov_2018} O.A.~Shcherbakov, A.S.~Vorobyev, E.M.~Ivanov, Phys. Part. Nucl. \textbf{49}, 81 (2018)

\bibitem{Tarrio_2014} D.~Tarrio, L.S.~Leong, L.~Audouin, I.~Duran, C.~Paradela et al. Nucl. Data Sheets {\bf 119}, 35 (2014)

\bibitem{Leal-Cidoncha_2016} E.~Leal-Cidoncha, I.~Duran, C.~Paradela, D.~Tarrio, L.S.~Leong et al. EPJ Web Conf. {\bf 111}, 10002 (2016)

\bibitem{Geppert-Kleinrath_2019} V.~Geppert-Kleinrath, F.~Tovesson, J.S.~Barrett, N.S.~Bowden, J.~Bundgaard et al. Phys. Rev. C {\bf 99}, 064619 (2019)

\bibitem{Ryzhov_2005} I.V.~Ryzhov, M.S.~Onegin, G.A.~Tutin, J.~Blomgren, N.~Olsson, A.V.~Prokofiev, P.-U.~Renberg, Nucl. Phys. A {\bf 760}, 19 (2005)

\bibitem{Bohr_1956} A.~Bohr, \textit{Proc. Int. Conf. on the Peaceful Uses of Atomic Energy, Geneva, 1955, v. 2} (UN, New York, 1956) p. 151

\bibitem{Vandenbosch_1973} R.~Vandenbosch, J.R.~Huizenga, \textit{Nuclear Fission} (Academic Press, New York, 1973) Chapters V and VI

\bibitem{Koning_2008} A.J.~Koning, S.~Hilaire, M.C.~Duijvestijn, \textit{Proc. Int. Conf. on Nuclear Data for Science and Technology, Nice, 2007} (EDP Sciences, 2008) p. 211

\bibitem{Capote_2009} R.~Capote, M.~Herman, P.~Oblozinsky, P.G.~Young , S.~Goriely et al. Nucl. Data Sheets {\bf 110}, 3107 (2009)

\bibitem{Shcherbakov_2002} O.~Shcherbakov, A.~Donets, A.~Evdokimov, A.~Fomichev, T.~Fukahori et al. J. Nucl. Sci. Tech. Suppl. {\bf 39 (suppl.~2)}, 230 (2002)

\bibitem{Paradela_2010} C.~Paradela, L.~Tassan-Got, L.~Audouin, B.~Berthier, I.~Duran et al. Phys. Rev. C {\bf 82}, 034601 (2010)

\bibitem{Diakaki_2016} M.~Diakaki, D.~Karadimos, R.~Vlastou, M.~Kokkoris, P.~Demetriou et al. Phys. Rev. C {\bf 93}, 034614 (2016)

\bibitem{Krappe_2012} H.J.~Krappe, K.~Pomorski, \textit{Theory of Nuclear Fission} (Springer, Heidelberg, 2012) Chapter 4

\bibitem{Blum_2012} K.~Blum, \textit{Density matrix theory and applications} (Springer, Heidelberg, 2012) Chapter 4

\bibitem{Brolley_1954} J.E.~Brolley,~Jr., W.C.~Dickinson, Phys. Rev. {\bf 94}, 640 (1954)

\bibitem{Simmons_1960} J.E.~Simmons, R.L.~Henkel, Phys. Rev. {\bf 120}, 198 (1960)

\bibitem{Leachman_1965} R.B.~Leachman, L.~Blumberg, Phys. Rev. {\bf 137}, B814 (1965)

\bibitem{Shpak_1971} D.L.~Shpak, B.I.~Fursov, G.N.~Smirenkin, Sov. J. Nucl. Phys. {\bf 12}, 19 (1971)

\bibitem{Iyer_1970} R.H.~Iyer, M.L.~Sagu, \textit{Proc. Nucl. and Solid State Physics Symp., Madurai, 1970, v. 2} (1970) p. 57

\bibitem{Androsenko_1982} Kh.D.~Androsenko, G.G.~Korolev, D.L.~Shpak, VANT Ser. Yadernye Konstanty {\bf 46(2)}, 9 (1982)

\bibitem{Ouichaoui_1988} S.~Ouichaoui, S.~Juhasz, M.~Varnagy, J.~Csikai, Acta Physica Hungarica {\bf 64}, 209 (1988)

\bibitem{Barabanov_1986} A.L.~Barabanov, D.P.~Grechukhin, Sov. J. Nucl. Phys. {\bf 43}, 892 (1986)

\bibitem{Cox_1953} J.A.M.Cox, H.A.Tolhoek, Physica {\bf 19}, 673 (1953)

\end{thebibliography}
\end{document}